\documentclass[sn-basic,Numbered]{sn-jnl}
\usepackage{graphicx}%
\usepackage{multirow}%
\usepackage{amsmath,amssymb,amsfonts}%
\usepackage{amsthm}%
\usepackage{mathrsfs}%
\usepackage[title]{appendix}%
\usepackage{xcolor}%
\usepackage{textcomp}%
\usepackage{manyfoot}%
\usepackage{booktabs}%
\usepackage{algorithm}%
\usepackage{algorithmicx}%
\usepackage{algpseudocode}%
\usepackage{listings}%
\usepackage{subcaption}%

\begin{document}

\title[Article Title]{MotifPiece: A Data-Driven Approach for Effective Motif Extraction and Molecular Representation Learning}

\author[1]{\fnm{Zhaoning} \sur{Yu}}\email{znyu@iastate.edu}


\author*[1]{\fnm{Hongyang} \sur{Gao}}\email{hygao@iastate.edu}

\affil*[1]{\orgdiv{Department of Computer Science}, \orgname{Iowa State University}, \orgaddress{\street{2434 Osborn Dr}, \city{Ames}, \postcode{50011}, \state{Iowa}, \country{United States}}}



\abstract{Motif extraction is an important task in motif based molecular representation learning. 
Previously, machine learning approaches employing either rule-based or string-based techniques to extract motifs. Rule-based approaches may extract motifs that aren't frequent or prevalent within the molecular data, which can lead to an incomplete understanding of essential structural patterns in molecules.
String-based methods often lose the topological information inherent in molecules. This can be a significant drawback because topology plays a vital role in defining the spatial arrangement and connectivity of atoms within a molecule, which can be critical for understanding its properties and behavior. 
In this paper, we develop a data-driven motif extraction technique known as MotifPiece, which employs statistical measures to define motifs. To comprehensively evaluate the effectiveness of MotifPiece, we introduce a heterogeneous learning module.
Our model shows an improvement compared to previously reported models. Additionally, we demonstrate that its performance can be further enhanced in two ways: first, by incorporating more data to aid in generating a richer motif vocabulary, and second, by merging multiple datasets that share enough motifs, allowing for cross-dataset learning.}

\keywords{Motif extraction, Molecular representation learning, Graph Neural Networks, Heterogeneous motif graph}



\maketitle

\section{Introduction}

Graph neural networks (GNNs) have showcased their proficiency in tackling various intricate tasks in the field of molecular property prediction. These tasks encompass activities such as classifying nodes within molecular graphs~\cite{kipf2016semi}, distinguishing between different molecular structures~\cite{xu2018powerful, song2020communicative, li2022graph}, and predicting intermolecular interactions~\cite{schlichtkrull2018modeling, huang2020skipgnn, wang2021multi}.  Instead of manually designing specific features, GNNs transform a molecular graph into a high-dimensional Euclidean space by leveraging the inherent topological relationships among its constituent nodes~\cite{scarselli2008graph}.

Traditional GNNs models primarily rely on the fundamental topology of molecular graphs to extract structural information~\cite{kipf2016semi, ying2018hierarchical, gao2019graph,liu2021spherical,wang2022advanced}. This is accomplished through techniques such as neighborhood feature aggregation and pooling methods. While these methods are powerful in capturing local structural features within individual molecular graphs, they typically do not explicitly focus on learning recurring motif patterns that span across different molecular graphs.

In molecular chemistry, a motif, also termed as a functional group, stands for a unique collection of atoms that are chemically bonded together in a consistent and repeating pattern~\cite{de2019synthetic, ertl2019systematic, ertl2020most, zhang2020motif, zhang2021motif}. This pattern can occur across various molecules. 
The distinguishing properties and chemical behavior of a molecule are significantly influenced by its motifs~\cite{debnath1991structure, yang2019surface, alizadeh2023influence}, as they represent the crucial reactive elements within the molecule. Over recent years, a growing body of scientific work has been dedicated to identify and incorporate the structural motifs in the learning of molecular representations~\cite{yu2022molecular, yang2022learning, zang2023hierarchical, fang2023single, zhang2020motif}.

The process of extracting motifs is an integral component in the motif-related study. The capability of a model to effectively gather valuable motif information mainly depends on how well this extraction process works. If the extraction isn't done correctly or accurately, a model might have trouble understanding and using the motif data effectively. Therefore, the model's success in handling motif information largely relies on how well the motif extraction process is executed. 
Previouos works employ either rule-based or string-based techniques to extract motifs. Rule-based approaches are based on domain knowledge and may extract motifs that aren't frequent or prevalent within the molecular data. This can lead to an incomplete understanding of essential structural patterns in molecules.
String-based methods use a string to represent a molecule and directly apply Natural Language Learning method to extract motifs. It may ignore the topological information inherent in molecules. This can be a significant drawback because topology plays a vital role in defining the spatial arrangement and connectivity of atoms within a molecule, which can be critical for understanding its properties and behavior. 

\begin{figure}[htbp]
\centering
\includegraphics[width=\textwidth]{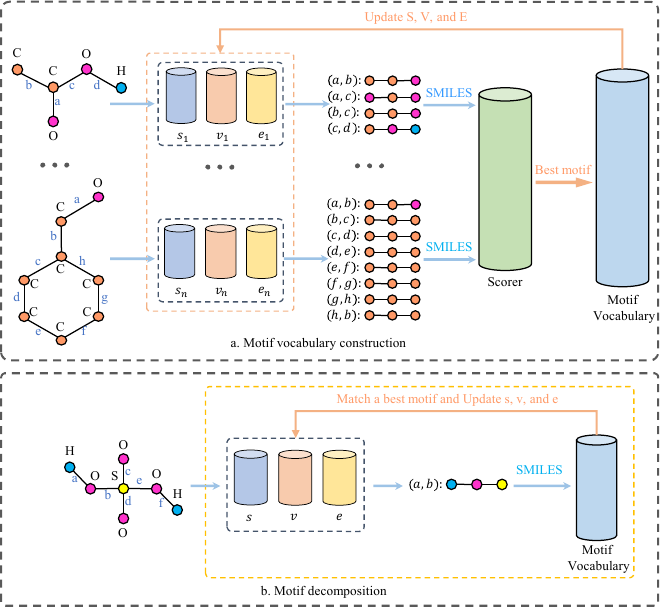}
\caption{\textbf{a. Motif vocabulary construction.} We show one iteration of a motif vocabulary construction. Given a molecule, we use three hash maps to represent it. Next, we consider all potential motif candidates, convert them into SMILES representations. After we've processed all the molecules, we assign a score to each motif candidate. We put the motif with the highest score into the motif vocabulary. Following this, the method update the three hash maps and proceed to the next iteration. \textbf{b. Motif decomposition.} We show one iteration of a motif decomposition. Again, we use three hash maps to represent a molecule. We begin by selecting a random node and creating a motif candidate. Then, we check if this motif candidate already exists in our motif vocabulary. If it does, we merge two edges that form the motif candidate and move on to the next motif candidate.}\label{fig:motifpiece}
\end{figure}

\begin{figure}[htbp]
\centering
\includegraphics[width=\textwidth]{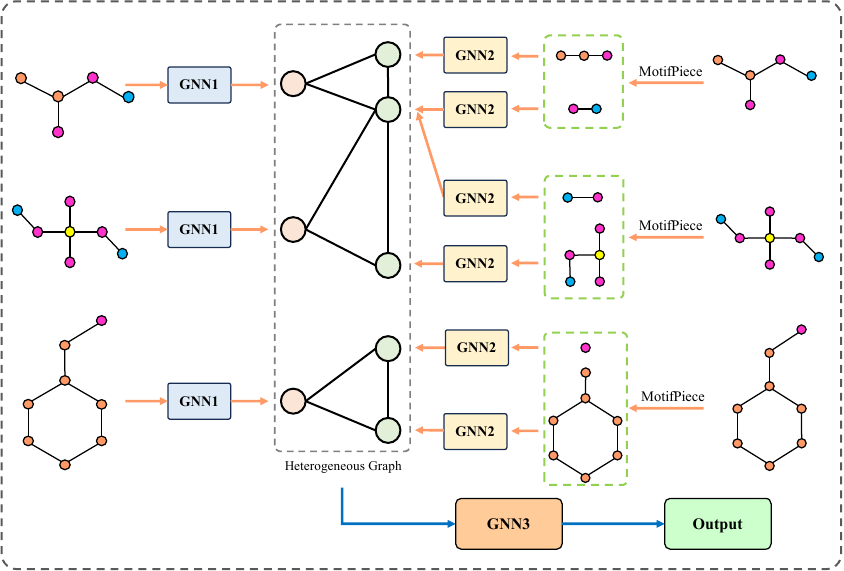}
\caption{\textbf{Heterogeneous Graph Learning Module.} The heterogeneous graph includes two distinct types of nodes: motif nodes and molecule nodes. The module employs two GNNs to extract graph embeddings for the motif graphs and molecule graphs. These embeddings serve as the features for their respective nodes. We then introduce a third GNN, which helps learn embeddings for every node within the heterogeneous graph.}\label{fig:hlm}
\end{figure}

\begin{figure}[htbp]
\centering
\includegraphics[width=\textwidth]{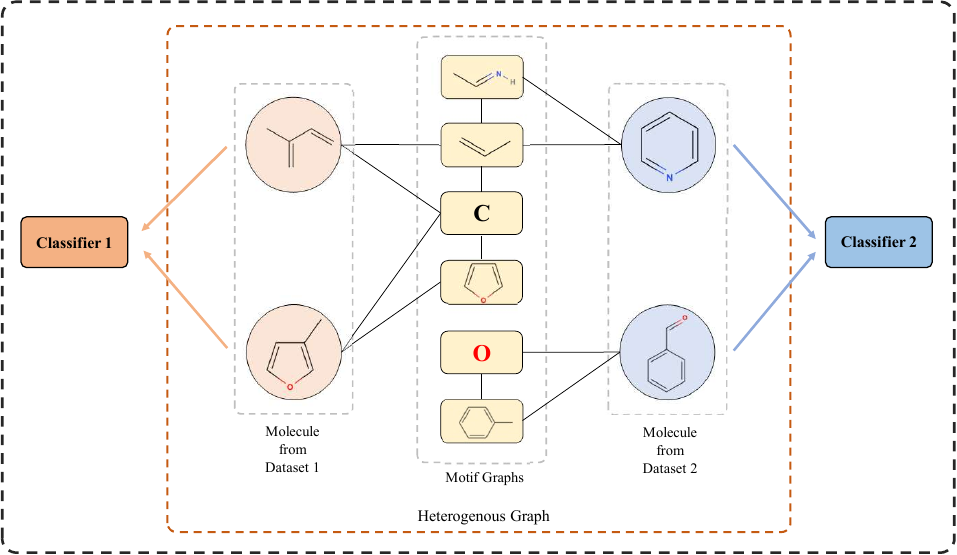}
\caption{\textbf{Cross Datasets Learning Module.} Here we show a simplified cross datasets learning module. 
In a heterogeneous graph containing two datasets, it has three types of nodes: motif nodes (all motifs from both datasets), molecule nodes from the first dataset, and molecule nodes from the second dataset. The node embeddings of the heterogeneous graph will be used for the downstream tasks.
If the task is graph classification, we introduce two separate classifiers, each tailored to one of the two datasets.}\label{fig:cdl}
\end{figure}

In this paper, we introduce a novel method called MotifPiece for extracting motifs from molecular data. MotifPiece is a data-driven approach that can adapt to the unique characteristics and patterns present in a group of molecules.
To comprehensively evaluate existing motif extraction methods and explore how motif information can enhance molecular representation learning, we propose a heterogeneous learning module. This module is designed to learn both motif embeddings and molecule embeddings within a heterogeneous graph. In this graph, motifs are represented as motif nodes, and molecules are represented as molecule nodes.
By leveraging the advantages offered by the heterogeneous graph, we integrate two types of embeddings: atom-level embeddings, and motif-level embeddings. This integration enhances the process of learning molecular representations, enabling a more holistic understanding of molecular structures and properties.
Furthermore, building upon the heterogeneous learning module, we discover that we can further enhance the learning of molecular representations through a cross-dataset learning module. This module allows us to combine different datasets by identifying shared motifs and training them together. This approach promotes the transfer of knowledge across diverse datasets and contributes to more robust and accurate molecular representation learning.
The pipeline of the overall framework is illustrated in Fig.~\ref{fig:motifpiece},  Fig.~\ref{fig:hlm}, and Fig.~\ref{fig:cdl}, which consists of the following modules:\\

\begin{enumerate}
    \item \textbf{MotifPiece}. The motif extraction module aims to decompose a molecule into recurring and statistically significant subgraphs. We introduce a data-driven motif extraction method called MotifPiece to identify underlying structural or functional patterns that are specific to the given set of molecules. As shown in Fig.~\ref{fig:motifpiece}, the MotifPiece module is comprised of two main parts. 
    \begin{enumerate}
        \item The first part, called motif vocabulary construction, is designed to create a vocabulary from a given set of molecules. This process involves iterating through all the molecules, and within each molecule, treating an edge with two nodes as a basic motif. We want to mention that the node here may contain several original nodes since it will be updated in each iteration. Any two connected basic motifs are combined to form a motif candidate. We then enumerate all possible motif candidates, keeping track of the frequency of each candidate for subsequent calculations. Once all the molecules have been iterated through, we have a collection of motif candidates. Each candidate is scored by frequency, and the motif with the highest score is selected and added to the motif vocabulary. Following this, we update the representation of the molecules (merge three nodes in the two basic motifs that form the selected motif into a single node). We continue to repeat this process, extracting additional motifs into the vocabulary until the score of all motif candidates falls below a predefined threshold.
        \item The second phase employs a greedy motif decomposition approach. During this stage, the objective is to break down a given molecule into multiple motifs. We begin by randomly selecting an initial edge, and from this starting point, we systematically iterate through all possible motif candidates. If any of these motif candidates match existing entries in a predefined motif vocabulary, which may have been generated by the first part of our model or defined from external sources, we merge the constituent nodes of the chosen motif candidate into a single node. We repeat this procedure until no more motif in the vocabulary can be identified within the molecule.
    \end{enumerate}
    \item \textbf{Heterogeneous Graph Learning}. From a given set of molecules, we employ MotifPiece to derive motifs. With these extracted motifs and the entire molecule set, we then construct a heterogeneous graph composed of both motif nodes and molecule nodes. Within this graph, each molecule node is linked to its associated motif nodes. Additionally, two motif nodes are connected if they share at least one identical atom in any molecule. For molecule representation learning, we initially deploy a learnable graph neural network to glean the atom-level embedding of the molecule, utilizing this embedding as the preliminary feature for the molecule node within the heterogeneous graph. Concurrently, a separate graph neural network is engaged to determine the embedding of motifs, serving as the initial feature for motif nodes. Ultimately, a third graph neural network is used to study the heterogeneous graph. The resulting embedding for the molecule node is then employed for subsequent operations, such as graph classification tasks.
    \item \textbf{Cross Datasets Learning}. By identifying shared motifs between two datasets, we can integrate them into a unified heterogeneous graph, allowing simultaneous learning of embeddings for both. When we apply MotifPiece to each set of molecules, we obtain two separate motif sets. These are then merged, resulting in a heterogeneous graph composed of motif nodes and two distinct sets of molecule nodes, each from one of the original datasets. To initialize, three distinct graph neural networks are employed: one for each node type in the graph, providing initial features for the molecule nodes from both datasets as well as for the motif nodes. Following this initial step, a fourth graph neural network is utilized to refine and learn the embeddings of all nodes within the heterogeneous structure. Finally, for downstream tasks, we deploy one classifier to process the molecule nodes from the first dataset and another for those from the second dataset.
\end{enumerate}

Analyzing motifs or functional groups is significant for molecule representation learning, and our approach achieves improvement over prior models. Our data-driven extraction method has the ability to reveal unique structural or functional patterns that are specific to a particular group of molecules. In heterogeneous motif learning module, we highlight how inter-molecular semantic motif information can enhance molecule representation learning. Additionally, merging datasets with a highly correlated motif vocabulary and jointly training them can further enhance performance.

The remainder of this paper is organized as follows: In the "Results" section, we presents results on graph classification tasks, highlighting the potency of our MotifPiece technique and showing outcomes from cross-dataset learning. The "Discussion" section delves into a comparative analysis, pinpointing the advantages and limitations of our methodology against existing models. Within the "Methods" section, we elaborate on our model's construction, encompassing MotifPiece, heterogeneous graph learning module, and cross-dataset learning module. We also discuss the time complexity of our MotifPiece approach.
\clearpage
\section{Results}

\subsection{Molecule classification results on individual dataset}

We first evaluate the effectiveness of our heterogeneous graph learning framework. We evaluate our model on ten widely-used datasets, four of them are from TUDataset~\cite{morris2020tudataset}, and others are from MoleculeNet dataset~\cite{wu2018moleculenet}. We compare our framework with three state-of-the-art baseline models: the Graph Isomorphism Network (GIN)~\cite{xu2018powerful}, the Higher-order Graph Neural Networks (HOGNN)~\cite{morris2019weisfeiler} and the Heterogeneous Motif Graph Neural Networks (HMGNN)~\cite{yu2022molecular}. We runs the datasets ten times, with both mean and standard deviations of results for the test set being reported.

\begin{table}[h]
\caption{Molecule classification results on indicidual dataset, the best performer on each dataset are shown in \textbf{bold}.}\label{tab: 2}
\begin{tabular*}{\textwidth}{@{\extracolsep\fill}lcccc}
\toprule%
& \multicolumn{4}{c}{TUDataset\footnotemark[1]} \\\cmidrule{2-5}%
 & PTC\_MR  & PTC\_MM  & PTC\_FR  & PTC\_FM \\
\midrule
GIN & 72.27 $\pm2.62$ & 76.51 $\pm1.55$ & 76.07 $\pm1.49$ &74.37 $\pm1.67$\\
HOGCN & 71.16 $\pm1.92$ & 75.48 $\pm1.37$ & 76.96 $\pm1.41$ & 72.59 $\pm1.98$\\
HMGNN & 73.95 $\pm0.95$  & 75.65 $\pm1.04$ & 77.70 $\pm1.23$ & 73.46 $\pm1.06$\\
Ours & 74.44 $\pm1.32$  & 78.76 $\pm1.04$ & 78.24 $\pm0.48$ & 75.41 $\pm1.47$\\
Ours (super vocab)  & \textbf{76.90 $\boldsymbol{\pm}$1.50} & \textbf{79.27 $\boldsymbol{\pm}$1.05}  & \textbf{79.06 $\boldsymbol{\pm}$1.89}  & \textbf{77.00 $\boldsymbol{\pm}$1.39} \\
\botrule
\end{tabular*}
\begin{tabular*}{\textwidth}{@{\extracolsep\fill}lccc}
\toprule%
& \multicolumn{3}{c}{MoleculeNet\footnotemark[2]} \\\cmidrule{2-4}%
 & BBBP  & SIDER  & CLINTOX   \\
\midrule
GIN   & 67.83 $\pm3.19$ & 59.59 $\pm1.82$  & 86.22 $\pm2.29$\\
HOGCN   & 67.35 $\pm1.43$ & 59.89 $\pm0.87$  & 78.00 $\pm3.68$\\
HMGNN & 68.81 $\pm3.89$ & 59.69 $\pm1.77$  & 85.02 $\pm5.15$\\ 
Ours  & \textbf{72.27 $\boldsymbol{\pm}$1.35} & \textbf{64.08 $\boldsymbol{\pm}$0.46}  & \textbf{88.98 $\boldsymbol{\pm}$3.16}\\
\botrule
\toprule%
 & BACE  & TOXCAST  & TOX21   \\
\midrule
GIN   & 76.94 $\pm1.21$ & 62.18 $\pm1.10$  & 74.29 $\pm0.59$\\
HOGCN   & 77.54 $\pm3.50$ & 62.32 $\pm0.69$  & 75.01 $\pm1.03$\\
HMGNN & 78.61 $\pm2.79$ & 63.12 $\pm0.96$  & 74.58 $\pm0.44$\\ 
Ours  & \textbf{81.08 $\boldsymbol{\pm}$2.51} & \textbf{64.85 $\boldsymbol{\pm}$0.65} & \textbf{75.80 $\boldsymbol{\pm}$0.88}\\
\botrule
\end{tabular*}
\footnotetext[1]{Note: We report accuracy for TUDataset.}
\footnotetext[2]{Note: We report AUC-ROC for MoleculeNet dataset.}
\end{table}


The evaluation results for the four datasets derived from the TUDataset are presented in the first part of Table~\ref{tab: 2}. When generating a motif vocabulary based on the dataset itself, the small size of all four datasets limits the full potential of our data-driven MotifPiece method. Notably, since these datasets all fall under the broad category of PTC, we conducted an additional comparison experiment: combining all four datasets to formulate a super motif vocabulary. This integration resulted in a performance surge of at least 1.36\% to 2.95\% when compared to other benchmarks. Furthermore, our combined motif vocabulary model saw a performance increment ranging from 0.51\% to 2.46\%. This observation underscores the ability of our heterogeneous learning framework to effectively represent smaller datasets. It also illuminates the significance of effective motif extraction for molecular representation learning and the benefits of augmenting relevant vocabulary, thereby enhancing motif extraction through our MotifPiece method.

On the other hand, the evaluation results for the six datasets from MoleculeNet are detailed in the second part of table~\ref{tab: 2}. In this context, our approach outperforms the three baselines, recording an AUC-ROC improvement between 0.79\% and 4.19\%. This reinforces the efficacy of our method in dealing with out-of-distribution scenarios. The notable performance boost in such settings is rooted in our method's design. Even with unique scaffolds spanning training, validation, and testing sets, a considerable number of common motifs persist across them. By acquiring motif-level embeddings, our approach is adept at navigating the intrinsic diversity of out-of-distribution situations, leading to marked enhancements in performance.

The advancements achieved by our method stem from three primary reasons: 1) Our MotifPiece method is quite effective at identifying important motifs, especially when it has access to a substantial amount of data. 2) While smaller datasets aren't ideal for data-driven approaches, the inclusion of supplemental vocabulary allows our MotifPiece to discern critical motifs, even within these smaller datasets, facilitating effective molecular representation learning.
3) The heterogeneous graph structure we employ connects molecules through common motifs, facilitating inter-molecular message passing. As shared motifs might exhibit analogous properties across varying molecules, the relay of this semantic information proves potent, augmenting the method's capacity to tackle out-of-distribution challenges.

\subsection{Study of Motif Extraction Methods}\label{sec:2.2}


In this section, we delve into a comprehensive analysis of the currently available motif extraction methods alongside our newly proposed method, MotifPiece. We conduct a series of intricate experiments to discern the strengths and weaknesses of each approach. By doing so, we aim to provide a constructive assessment that could serve as a valuable guide for future research endeavors.

We compare our method with three widely used extraction methods. The first one is the well-established decomposition method, for which we have elected to use the BRICS algorithm~\cite{degen2008art} as a representative. BRICS is a method that identifies and breaks rotatable bonds in a molecule, resulting in smaller, manageable chemical substructures. Secondly, we draw a comparison to a method that focuses on the severance of bridge bonds to extract molecular motifs. A bridge bond connects two parts of a molecule that would otherwise form two distinct entities~\cite{jin2020hierarchical}. The third method we consider involves the extraction of simple ring structures and their connecting bonds as motifs (R\&D)~\cite{jin2018junction}. This approach focuses on identifying and isolating simple cyclic structures within a molecule, which often play key roles in the molecule's chemical behavior.

We have also introduced two modified variants of the MotifPiece method, each adopting a distinct decomposition strategy. In contrast to the original approach, which employs a greedy search for motif candidates, these two variants take a different path. They begin by generating all potential candidates and subsequently select the one with the highest score based on a predefined scoring function.
In our experimental configuration, we apply two scoring methods. The first one aligns with the score method used during motif vocabulary generation, which assesses the frequency of a motif candidate (Variant 1). The second method evaluates the number of atoms within a motif candidate (Variant 2). These two heuristic variants enable a more comprehensive consideration of the motif candidate's characteristics on a global scale.

\begin{table}[h]
\caption{Motif Extraction Study on MoleculeNet Datasets, the best performer on each dataset are shown in \textbf{bold}.}\label{tab: motifpiece}
\begin{tabular*}{\textwidth}{@{\extracolsep\fill}lcccc}
\end{tabular*}
\begin{tabular*}{\textwidth}{@{\extracolsep\fill}lccc}
\toprule%
& \multicolumn{3}{c}{MoleculeNet\footnotemark[1]} \\\cmidrule{2-4}%
 & BBBP  & SIDER  & CLINTOX   \\
\midrule
BRICS   & 66.03 $\pm1.51$ & 59.19 $\pm1.94$  & 86.52 $\pm3.11$\\
Bridge   & 63.93 $\pm1.72$ & 58.23 $\pm2.10$  & 86.96 $\pm2.98$\\
R\&B & 69.82 $\pm1.54$ & 59.98 $\pm2.02$  & 84.81 $\pm2.23$\\ 
Variant\_1 & 69.90 $\pm1.21$ & \textbf{65.12 $\boldsymbol{\pm}0.72$} & 84.65 $\pm3.33$\\
Variant\_2  & 71.12 $\pm 1.28$ & 64.97 $\pm$ 0.94 & 87.39 $\pm 0.30$\\
MotifPiece  & \textbf{72.27 $\boldsymbol{\pm}$1.35} & 64.08 $\pm$0.46  & \textbf{88.98 $\boldsymbol{\pm}$3.16}\\
\botrule

\toprule%
& \multicolumn{3}{c}{MoleculeNet\footnotemark[1]} \\\cmidrule{2-4}%
 & BACE  & TOXCAST  & TOX21   \\
\midrule
BRICS   & 80.64 $\pm$1.00 & 63.95 $\pm0.52$  & 75.12 $\pm0.66$\\
Bridge   & 78.05 $\pm1.65$ & 61.39 $\pm1.26$  & 73.89 $\pm1.22$\\
R\&B & 78.86 $\pm1.99$ & 63.65 $\pm0.72$  & 75.03 $\pm0.65$\\ 
Variant\_1 & 77.13 $\pm 2.23$ & 64.47 $\pm 0.57$ & 75.13 $\pm 0.57$\\
Variant\_2 & 77.23 $\pm 1.14$ & 64.69 $\pm 0.91$ & 75.06 $\pm 1.24$\\
MotifPiece  & \textbf{81.08 $\boldsymbol{\pm}$2.51} & \textbf{64.85 $\boldsymbol{\pm}$0.65} & \textbf{75.80 $\boldsymbol{\pm}$0.88}\\

\botrule
\end{tabular*}
\footnotetext[1]{Note: We report AUC-ROC for MoleculeNet datasets.}
\end{table}

We evaluated all methods across the six MoleculeNet datasets mentioned earlier. To ensure a fair comparison, we assessed all extraction methods within the framework of our heterogeneous learning system. We maintained consistent model architectures, including the number of layer, and batch normalization throughout.


The table~\ref{tab: motifpiece} shows the results for the six MoleculeNet datasets. Here, we evaluate the performance based on AUC-ROC metric. Similar to the previous test, we runs the datasets ten times, with both mean and standard deviations of AUC-ROC for the test set being reported. Our MotifPiece consistently outperforms three widely used competitors, showing a lead ranging between 0.44\% and 4.10\%. It demonstrates the effectiveness of our data-driven method. 

We claim that the suboptimal performance observed in the extraction of simple rings and bonds may stem from the inherent lack of complexity in these motifs, thereby posing challenges for effective processing within the heterogeneous learning module. Regarding the BRICS and Bridge methods, which rely on knowledge-guided extraction, their underperformance may be attributed to their consistent extraction of intricate and infrequent motifs. This persistent extraction of complex and less common motifs could potentially hinder the meaningful information exchange between distinct molecules within the heterogeneous graph. In contrast, our approach leverages the strengths of both R\&B and knowledge-based extraction, accomplishing the dual goal of increasing the intricacy of each motif while preserving its frequency.

We also conducted a comparative analysis between our greedy-based decomposition method and two heuristic decomposition method (Variant\_1 \& Variant\_2). The results indicate that the greedy-based method outperforms the heuristic decomposition method on 5 out of 6 datasets. We suggest that the results obtained from Variant\_1, which employs motif decomposition based on frequency as the score function, might be influenced by variations in the frequency distribution between the training and test sets. Focusing more on motif frequency can potentially bias motif extraction towards the distribution in the training set.
As for the outcomes of Variant\_2, which utilizes the number of atoms in a molecule as the score function, they may be attributed to an excessive emphasis on motif size. 
While we generally prefer finding bigger motifs in motif extraction, this strong emphasis on size can sometimes lead us to accidentally include smaller, less important parts, especially when we're dealing with exceptionally large motifs.
On the contrary, our greedy decomposition method adopts an approach that treats all motifs within the vocabulary with equal importance. It also avoid an excessive obsession on motif size. This balanced perspective allows us to effectively extract motifs that strike a reasonable balance between frequency and size, ensuring that the resulting motifs are both meaningful and suitable for the task at hand.

The enhanced performance of our MotifPiece approach arises from two key aspects: 1) While MotifPiece is inherently data-driven, enhancing the motif vocabulary can help motif extraction on small datasets. In our tests, amalgamating the four TUDataset proved effective in expanding this vocabulary. Given the inherent relationship and relevance among these TUDatasets, this approach yielded promising results. 2) For datasets with a sufficient number of molecules, our method is adept at discerning and extracting informative and meaningful motifs, which are pivotal for efficient molecule representation learning. 3) The greedy decomposition method find a good balance between size of a motif and the frequency of the motif. This makes sure that the motifs it identifies are not too big or too rare, resulting in meaningful and suitable motifs.
\begin{figure}[htbp]
\centering
\includegraphics[width=\textwidth]{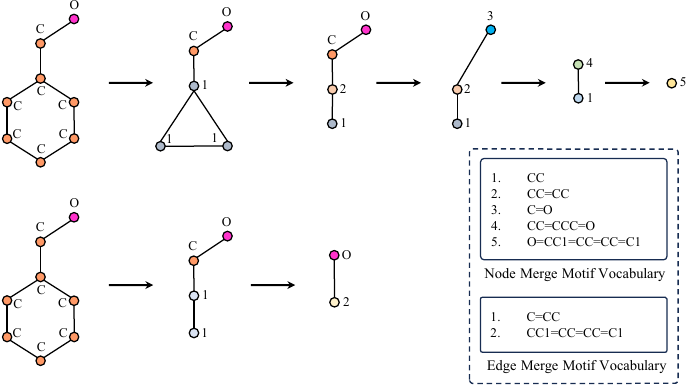}
\caption{An example of different merge methods}\label{fig:Merge_method}
\end{figure}

\subsection{Study of Merge Methods in MotifPiece}

In this section, we examine two distinct merging approaches within our MotifPiece framework. The first approach, known as "node merge," involves considering each basic motif as an individual node. Under this method, the algorithm merges a pair of nodes into a motif during each iteration. The second approach, termed "edge merge," defines a basic motif as an edge with two associated nodes. In this case, the method combines a pair of edges, effectively involving three nodes, to form a motif in each iteration.

As illustrated in Figure~\ref{fig:Merge_method}, when employing node merge, the example molecule \texttt{O=CC1=CC=CC=C1} can undergo a maximum of five merging operations, whereas with edge merge, only two merge operations are necessary. Node merge results in a motif vocabulary comprising five distinct motifs, while edge merge yields a vocabulary of just two motifs. Consequently, it can be inferred that the node merge strategy explores a more extensive motif candidate space, whereas the edge merge approach might ignore certain intermediate motif candidates. Simultaneously, the node merge method may encounter challenges in attaining global optima, whereas the edge merge approach excels in extracting motifs from a broader and more comprehensive perspective.

\begin{figure}[htbp]
\centering
\includegraphics[width=\textwidth]{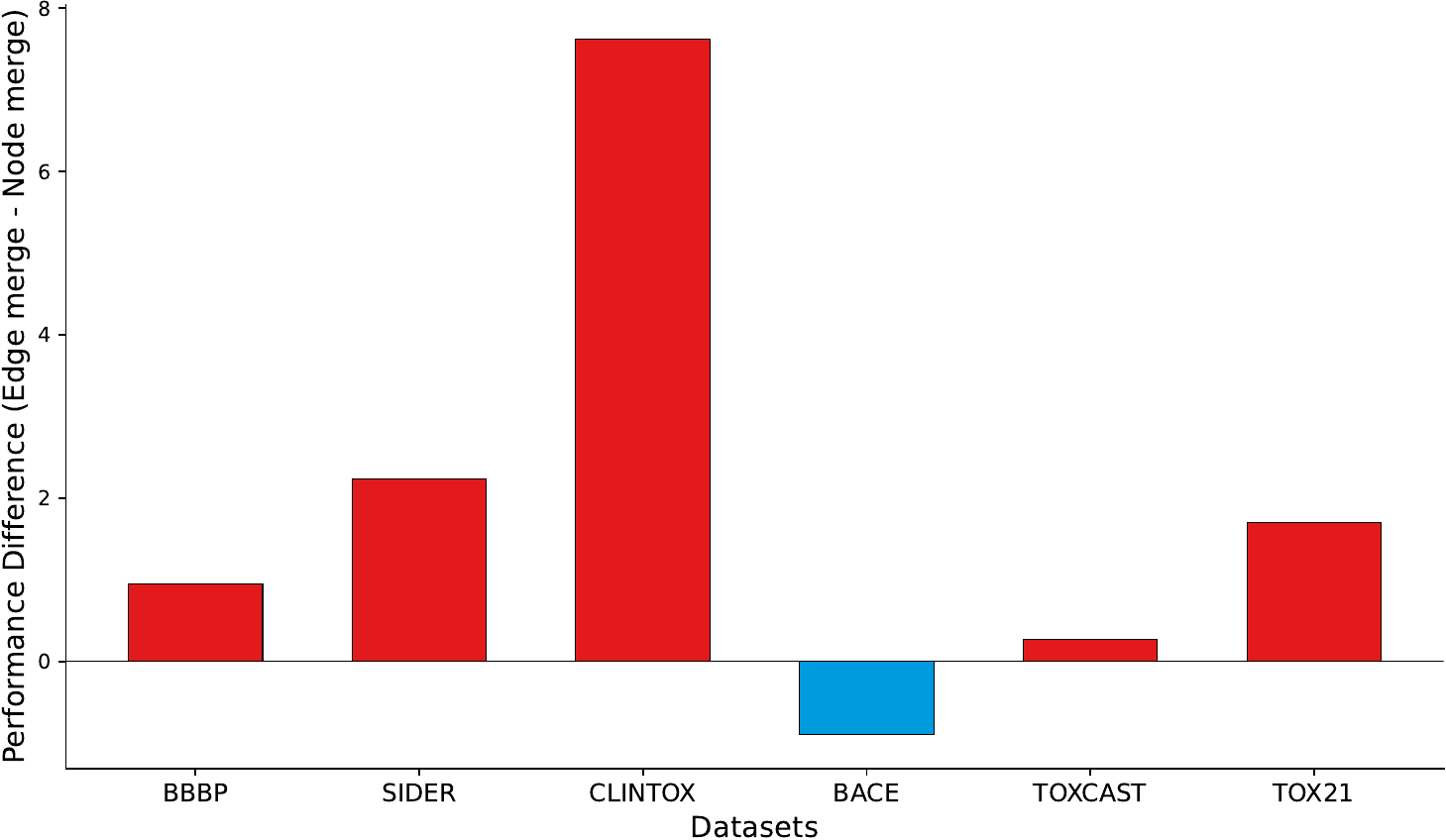}
\caption{Performance difference of different merge methods}\label{fig:merge_compare}
\end{figure}

\begin{figure}[htbp]
\centering
\includegraphics[width=\textwidth]{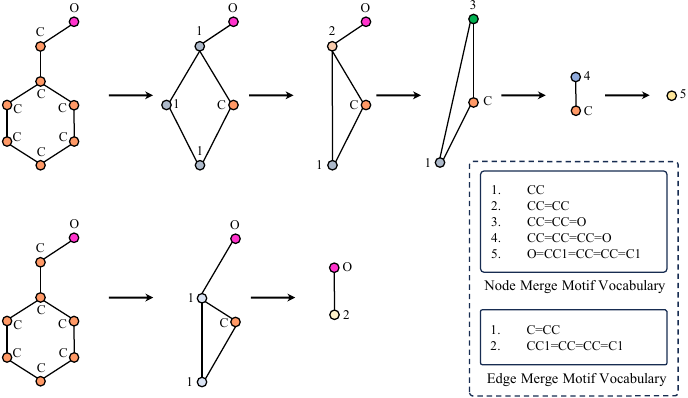}
\caption{Another example of different merge methods}\label{fig:Merge_method_2}
\end{figure}


In Figure~\ref{fig:merge_compare}, it is evident that, although node merge can identify a greater number of potential motif candidates, edge merge MotifPiece outperforms node merge MotifPiece in 5 out of 6 datasets. This result can be explained by the fact that the selection of which pairs of basic motifs to merge has a big impact on the later merging steps.

As illustrated in Figure~\ref{fig:Merge_method}, consider the initial merging step where we aim to combine two \texttt{C} nodes to create a \texttt{CC} motif. However, within the molecule, there are multiple occurrences of \texttt{CC}, and due to the absence of a prescribed order in the graph structure, there are no clear rules to guide us in selecting which node pairs to merge. This inherent uncertainty is magnified as the number of merge steps increases, ultimately impacting the composition of our motif vocabulary.

We can further illustrate this phenomenon in Figure~\ref{fig:Merge_method_2}, where different pairs of \texttt{C} nodes are selected for merging in the first iteration. It is apparent from the results that node merge yields a motif vocabulary with two distinct motifs, in contrast to the previous example, while edge merge maintains the same motif vocabulary as before. We contend that this discrepancy arises because node merge involves a greater number of merge iterations than edge merge, introducing heightened uncertainty throughout the process. This increased uncertainty compromises the consistency of motif extraction.

\subsection{Cross datasets performance evaluation}
In section~\ref{sec:2.2}, we observed that the performance of our heterogeneous learning module is limited when applied to smaller datasets. This constraint is likely due to the scarcity of semantic information available for the model to harness effectively. To mitigate this limitation and unlock its full potential, we explored alternative strategies. Beyond simply employing an augmented motif vocabulary, we also experimented with an approach where we amalgamate two datasets and jointly train them. This combined training not only introduces more variability and depth to the data but also allows for richer semantic interplay, which could potentially enhance the model's ability to capture intricate patterns and relationships within molecules.
\subsubsection{Dataset analysis}
\begin{figure}[htbp]
\centering
\includegraphics[width=0.7\textwidth]{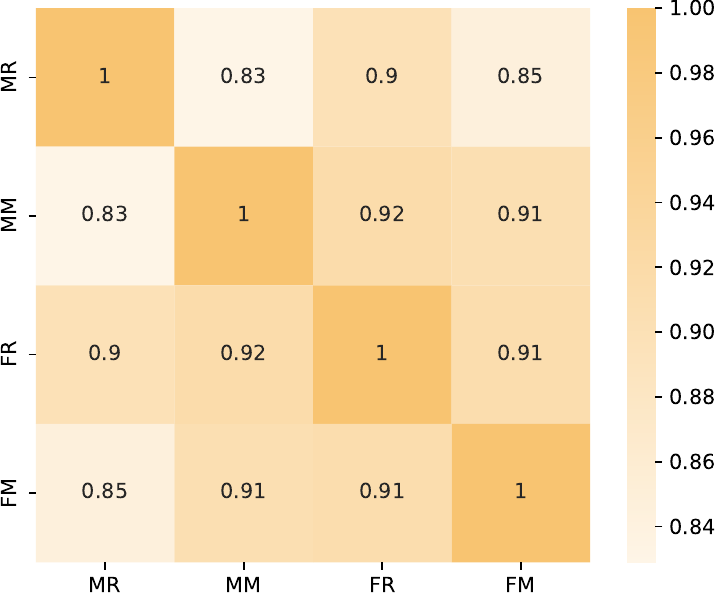}
\caption{Visualization of Jaccard Index values among four TUDatasets}\label{fig:JI}
\end{figure}
It is essential to note that the choice of merged datasets cannot be made arbitrarily. The crux of our approach hinges on the utilization of common motifs found across different datasets, amalgamating them into a heterogeneous graph.
If multiple datasets have limited overlap in motifs, their combination might not enhance the learning of molecular representations. In fact, it could potentially introduce noise into the heterogeneous graph, thereby hindering its overall effectiveness. To achieve optimal results with the combined dataset, it's desirable that the two datasets share a substantial number of common motifs.

To analyze whether the TUDatasets can be combined to train them together, we use a widely-used measure of similarity for sets - the Jaccard Index to measure the similarity between motif vocabularies for the TUDatasets. The Jaccard Index, also referred to as the Jaccard similarity coefficient, is computed as the size of the intersection divided by the size of the union of the sample sets. Mathematically, if we denote two sets as A and B, the Jaccard Index is defined as:
\begin{equation}
    J(A, B) = \frac{|A \cap B|}{|A \cup B|}
\end{equation} 
In our case, a higher Jaccard Index indicates a greater overlap of motifs between the datasets, whereas a lower index signifies fewer common motifs.
This measure assists us in grouping the datasets based on their degree of motif overlap, thereby providing an informed basis for our experimental design and subsequent analysis. By grouping datasets with Jaccard indices, we can explore our method's performance under different motif commonality levels.

The four TUDatasets - PTC\_MR, PTC\_MM, PTC\_FR, and PTC\_FM - all belong to the broader PTC dataset category. A notable feature among these datasets is their significant motif overlap, as evidenced by their high Jaccard indices in Fig.~\ref{fig:JI}. This overlap highlights the deeply intertwined nature of their content.

Our investigation into these datasets serves a dual purpose. Firstly, we want to spotlight the efficacy of our novel cross-dataset heterogeneous learning approach, emphasizing its adaptability across datasets with shared motifs. Secondly, we aim to validate the idea that when datasets have a high degree of shared motifs, combining them can enhance the quality of molecular representation learning. In essence, the shared patterns between these datasets can be leveraged to make the learning process more thorough and precise.

\begin{sidewaystable}
\caption{Graph classification accuracy (\%) on the combination of four TUDatasets}\label{tab3}
\begin{tabular*}{\textheight}{@{\extracolsep\fill}lcccccccc}
\toprule%
& \multicolumn{4}{c}{PTC\_MR}& \multicolumn{4}{c}{PTC\_MM}  \\\cmidrule{2-5}\cmidrule{6-9} %
Methods &  & + PTC\_MM & + PTC\_FR & + PTC\_FM	&  & + PTC\_MR & + PTC\_FR & + PTC\_FM \\
\midrule
Ours   & 76.90 $\pm1.50$ & 78.7 $\pm1.07$  & 81.6 $\pm1.65$  & 78.05 $\pm0.82$ & 79.27 $\pm1.05$ & 80.45 $\pm1.09$ & 80.96 $\pm2.02$ & 83.62 $\pm1.67$   \\
Accuracy change & 0.00 & +1.80 & +4.70 & +1.15 & 0.00 & +1.18 & +1.69 & +4.35 \\
Jaccard Index & -  & 0.8289 & 0.9061 & 0.9162 & - & 0.8289 & 0.8451 & 0.8980 \\
\botrule
\toprule
&  \multicolumn{4}{c}{PTC\_FR} & \multicolumn{4}{c}{PTC\_FM}  \\\cmidrule{2-5}\cmidrule{6-9} 
Methods &  & + PTC\_MR & + PTC\_MM & + PTC\_FM &  & + PTC\_MR & + PTC\_MM & + PTC\_FR \\
\midrule
Ours  & 79.06 $\pm1.89$ & 85.56 $\pm1.33$ & 80.22 $\pm2.06$ & 81.75 $\pm1.13$ & 77.00 $\pm1.39$ & 79.80 $\pm0.90$ & 81.97 $\pm1.12$ & 80.34 $\pm1.27$ \\
Accuracy change & 0.00 & +6.50 & +1.16 & +2.69 & 0.00 & +2.80 & +4.97 & +3.34 \\
Jaccard Index & -  & 0.9061 & 0.8451 & 0.9121 & - & 0.9162 & 0.8980 & 0.9121 \\
\botrule

\end{tabular*}
\end{sidewaystable}

\subsubsection{Molecule classification results}
Expanding upon the heterogeneous learning approach discussed in the previous section, we employed a 10-fold cross-validation technique to evaluate the efficacy of our integrated datasets within the heterogeneous learning paradigm. From our four datasets, we paired each dataset with another, resulting in combined sets for joint training sessions.

Upon examining the results, it's evident that there's a significant increase in accuracy when we combine datasets and train them jointly, in comparison to training them individually. Specifically, PTC\_MR's performance see an uplift ranging from 1.15\% to 4.70\%, PTC\_MM improves by 1.18\% to 4.35\%, PTC\_FR shows an enhancement from 1.16\% to 6.50\%, and PTC\_FM witnesses an increase between 2.80\% to 4.97\%. This boost in performance underscores the advantages of integrating two smaller datasets with a high Jaccard Index, leading to a richer set of training samples and a more dynamic semantic interchange.
A notable insight from our experiments is the clear link between the Jaccard Index of two datasets' motif vocabularies and the subsequent performance boost. Specifically, motif vocabulary pairs with a Jaccard Index surpassing 85\% displayed especially favorable outcomes. For these pairings, the accuracy improvement ranged from 2.69\% to as much as 6.5\%, highlighting the pivotal role of motif commonality in enhancing performance. Conversely, when the Jaccard Index was below 85\%, the increase in accuracy remained below 1.8\%.

The observations from this experiment can be ascribed to two primary reasons: 1) Datasets with a high Jaccard Index for motif vocabularies are rich in semantic content. Within the context of our cross-dataset heterogeneous learning model, this valuable information becomes particularly salient. When such datasets are trained simultaneously, our model capitalizes on this shared knowledge, leading to marked performance improvements for both datasets. 2) An greater Jaccard Index signifies a higher degree of motif commonality between datasets. This overlap not only fosters a deeper understanding but also empowers the model to craft more comprehensive and precise molecule representations. The symbiotic relationship between these shared motifs aids in refining the nuances of molecular characteristics across the combined datasets.

\clearpage
\section{Discussion}
In this section, we address the advantages and disadvantages of our approach when compared with other existing approaches, which are from each three part of our framework.

\subsection{MotifPiece}
There are three commonly employed methods for motif extraction:
1) Ring and Bonded Pairs (R\&B): This method focuses on extracting fundamental rings and bonds as motifs. While it effectively captures essential ring structures, it tends to overlook motifs that aren't strictly rings or bonds.
2) Molecule Decomposition: In this category, methods decompose molecules based on domain-specific knowledge. While they can uncover chemically significant motifs, the motifs they identify may not be prevalent in the dataset. Additionally, this domain knowledge is often generic and doesn't offer dataset-specific information.
3) Molecule Tokenization: This method leverages subword techniques inspired by Natural Language Processing (NLP) to extract motifs from the string representations of molecules. It is user-friendly and capable of extracting motifs from a comprehensive dataset perspective since it accesses the entire dataset for extraction. However, representing a graph as a string introduces identifiers like brackets, asterisks, and numbers to preserve topology information. Existing methods tend to overlook these identifiers during motif extraction, resulting in the loss of crucial topology information from the graph.

In contrast to other existing approaches, our method offers several noteworthy advantages, primarily stemming from its data-driven nature:
1) Flexibility in Motif Size: Our approach possesses the flexibility to identify motifs of varying sizes, provided they meet specific measurement criteria defined within our method. This adaptability allows us to capture motifs ranging from small and intricate structures to larger, more complex patterns.
2) Statistical Measures: Our algorithm incorporates statistical measures as a fundamental component of motif extraction. This statistical approach ensures that any identified motif is inherently common within the dataset. This means that the motifs identified are not mere outliers or anomalies but represent prevalent and meaningful patterns.
3) Graph-Level Extraction: One of our key strengths lies in the extraction of motifs at the graph level. This approach retains the complete topological information of the graph, preserving the intricate relationships and connections between nodes and edges. Consequently, our method excels in maintaining the structural integrity of the underlying graph while identifying motifs, contributing to a more comprehensive understanding of the data.

The primary limitation of our approach is also its data-dependent nature, which necessitates a sufficient volume of data to yield robust results. When working with relatively small datasets, our MotifPiece algorithm may struggle to achieve promising performance levels. In such cases, additional datasets become essential to enhance motif extraction outcomes.
In our experiments, we addressed this limitation by combine four PTC datasets, thereby creating an extensive and comprehensive super motif vocabulary. This strategic had a profound impact on the performance of our method, leading to a significant improvement in its effectiveness. 
Additionally, when conducting cross-dataset learning using multiple small datasets that share a motif vocabulary with high Jaccard Index, it can enhance overall performance.

\subsection{Heterogeneous Learning Module}

Our approach, which leverages a heterogeneous graph structure based on shared motifs for molecule representation learning, offers several advantages over traditional message-passing GNNs. 1) Traditional GNNs typically rely on neighborhood aggregation to learn node embeddings. While effective, this approach might not capture higher-level structural patterns and motifs within the graph. Our method, on the other hand, explicitly incorporates shared motifs, which are recurring subgraph patterns in molecular structures. This allows the model to capture important structural and chemical features that can significantly influence molecular properties. 2) One of the unique strengths of the approach is the ability to perform inter-molecular message passing. In traditional GNNs, messages are typically passed only within the same molecule, limiting their capacity to capture interactions and dependencies between different molecules. By allowing information to flow seamlessly between molecules based on shared motifs, our method can capture the influence of neighboring molecules, which is vital for understanding molecular interactions and predicting properties in complex chemical systems. 3) Shared motifs often carry semantic information about molecular structures, functions, or properties. Our approach leverages this semantic information effectively by propagating it through the graph. This means that the model can learn not only from the immediate neighborhood of a node but also from the structural similarities between molecules. This semantic transmission enhances the model's ability to generalize across different molecules and address out-of-distribution challenges, making it more robust and versatile.

Our approach's primary constraint lies in its omission of an explicit mechanism for learning interactions between motifs, potentially resulting in a limited understanding of motif interactions. Nonetheless, devising a method for explicitly learning motif interactions is a complex endeavor that poses substantial challenges. As such, this aspect remains an avenue for future research and development.

\subsection{Cross Datasets Learning}
Traditional graph neural networks are typically designed to learn from a single dataset at a time. In contrast, our approach offers a unique capability: it can integrate and learn from multiple datasets simultaneously by leveraging shared motifs. This ability to combine diverse datasets provides several advantages and opens up new possibilities for more comprehensive and robust learning. 1) Traditional graph neural networks can only handle one dataset at a time. This limitation makes it hard to use knowledge from different sources together. Our method, however, doesn't have this restriction. It lets us combine various datasets into a single learning system. This is particularly beneficial when dealing with related data sources, such as molecular structures from different chemical databases or social networks from various online platforms. 2)  By jointly learning from multiple datasets, our method can capture common structural and semantic patterns that are shared across diverse data sources. This enhanced generalization is particularly valuable in scenarios where each dataset may contain unique insights or represent different aspects of a complex problem. 3) Real-world data can be noisy and exhibit variations across different datasets. Learning from multiple datasets can help mitigate the impact of dataset-specific biases or outliers, leading to more stable and reliable model performance.

The main limitation of the cross datasets learning module is that the module's effectiveness hinges on the existence of common structural patterns or motifs across the datasets being integrated. These shared motifs serve as the foundation for knowledge transfer and harmonious learning. Without an adequate number of shared motifs, the module may struggle to find meaningful connections between the datasets. It can also lead to conflicting or confusing information. As a result, the model may produce less accurate predictions or even generate erroneous insights.

\clearpage
\section{Methods}

\subsection{Motif Extraction}
Motifs, which inherently correlate with molecular properties \cite{debnath1991structure}, significantly influence the performance of molecular graph learning models. Integrating motifs into graph learning requires motif identification and decomposition from molecules.
\textit{Motif identification} refers to recognizing motifs from molecules based on structural or chemical properties. The recognized motifs form a motif vocabulary. On the other hand, \textit{motif decomposition} involves breaking down a given molecule into motifs leveraging the motif vocabulary established during the identification stage.



\subsubsection{Existing methods}
Currently, motif extraction can be achieved through three primary methodologies.
1) Ring and bonded pairs (R\&B). Methods~\cite{jin2018junction, yu2022molecular, bouritsas2022improving} in this category recognize simple rings and bonded pairs as motifs.
2) Molecule decomposition. In this category, methods recognize motifs through molecule breakdown. These methods initially establish a catalog of target bonds, either through domain-specific knowledge, such as the ``breakable bonds'' used by RECAP \cite{zhang2021motif} and BRICS \cite{zang2023hierarchical,jiang2023pharmacophoric}, or through hand-crafted rules, like ``bridge bonds'' \cite{jin2020hierarchical}. The algorithms remove these target bonds from the molecules and use the consequent molecular fragments as motifs. 
3) Molecule tokenization. Certain methods~\cite{fang2023single,kuenneth2023polybert} leverage the string representations of molecules, such as SMILES, and directly apply the WordPiece algorithm from the NLP domain. The resulting strings within the constructed vocabulary are regarded as motifs. While these motif strings are adaptive to molecular data, treating molecules purely as strings could introduce problems such as the loss of chemical semantics.

\subsubsection{Data-driven motif vocabulary construction}\label{Task:3.1.2}

\begin{algorithm}
\caption{Function \textproc{Initialize()}}\label{alg: Initial}
\begin{algorithmic}[1]
\State {\bfseries Input:} Graph $G$
\State {\bfseries Output:} Subgraph hash map $S$, Node hash map $V$, Edge hash map $E$
\State {\bfseries Initialization:} Three empty hash map $S$, $V$, $E$

\For{each node $v$ in $G$}
    \State Add key-value pair $(v, v)$ to $V$
\EndFor

\For{each edge $e = (u, v)$ in $G$}
    \State Add key-value pair $(e, (u, v))$ to $E$

    \For{$j$ in $(u, v)$}
        \If{$j$ not in $S$}
            \State Add key-value pair $(j, \{e\})$ to $S$
        \Else
            \State Add $e$ to the $S[j]$
        \EndIf
    \EndFor
\EndFor
\end{algorithmic}
\end{algorithm}

We propose a data-driven approach, which applies statistical measures to define motifs. The proposed approach will employ a bottom-up algorithm that starts from simpler motifs (bonded pairs) and progressively assembles more complex ones.

In the initialization stage, we will construct a \textit{motif vocabulary} to store recognized motifs.
For each graph, the algorithm employs the \textproc{Initialize()} function to establish three hash maps, denoted as $S$, $V$, and $E$, for storing graph-related information. Specifically, $S$ functions as a mapping that links a node to a set of edges, where each edge in the set includes the node. $V$ serves as a mapping that correlates each node with its original node set. Lastly, $E$ represents a mapping that associates an edge with the two nodes that constitute it. Algorithm~\ref{alg: Initial} shows the details of \textproc{Initialize()} function.

Throughout the algorithm's iterations, these hash maps—$S$, $V$, and $E$—for each molecule will be consistently maintained and updated. This involves the replacement of simpler motifs with more complex ones.

Upon initialization, the algorithm will enter a phase of iteration, wherein a new motif will be added to the vocabulary in each cycle. The steps will be as follows: \\
1) Motif candidate enumeration: Each molecule will enumerate subgraph hash map $S$. For each key node in $S$, the value is an edge set where any edge in the set contains the key node. Thus, any two edges in the set can be seen as a connectable motif pair. A motif pair is considered connectable if they share the same atoms in the original molecule. Each pair will then be merged into a motif candidate, containing all the atoms and bonds from both motifs. The unique motif candidates across all molecules will form the motif candidate list. \\
2) Candidate frequency counting: For each motif candidate, we will calculate its occurrence frequency across all molecules. \\
3) New motif recognition: If the frequency of the most common motif candidate surpasses a threshold, it will be added to the motif vocabulary as a new motif. \\
4) Hash maps update: The new motif will be added to the hash maps in which it appears. In specific, if a molecule contains the new motif which contains three nodes, we first merge three nodes that form the motif into one node in hash map $V$. Then, we delete these three nodes in hash map $S$, and add the new node into $S$. We also modify nodes in $S$ that have any edges with any one of these three nodes. At last, those modified edges will be updated in hash map $E$.
The newly identified motif will be incorporated into the respective hash maps where it is found. To elaborate, when a molecule contains the newly discovered motif consisting of three nodes, the following steps are undertaken: First, the three nodes composing the motif are consolidated into a single node within the hash map labeled as $V$. Then, the aforementioned three nodes are removed from the hash map $S$, and the new node is introduced into $S$. After that, nodes within $S$ that are connected to any of these three nodes are adjusted accordingly. Subsequently, the modified edges are updated in the hash map $E$.

The algorithm will continue until the motif vocabulary reaches a pre-specified size or no new motifs are identified. Eventually, a motif vocabulary of motifs unique to the given molecular data will be generated. This process is described in Algorithm~\ref{alg:motif-defining}.

\begin{algorithm}
    \caption{Data-driven motif vocabulary construction}
    \label{alg:motif-defining}
    \begin{algorithmic}
    \State {\bfseries Input:} Molecules $\mathcal{G} = \{G_1, \dots, G_n\}$, vocabulary size $V$
    \State {\bfseries Output:} Motif vocabulary $\mathcal{M}$.
    \State {\bfseries Initialization:} Motif vocabulary $\mathcal{M}$ with distinct bonded pairs in $\mathcal{G}$. 
    
    \While{$|\mathcal{M}| < V$}
    \For{$G_i$ in $\mathcal{G}$}
        \If{$|\mathcal{M}| == 0$}
            \State $S_i$, $V_i$, $E_i$ $\Leftarrow$ $\textproc{Initialize($G_i$)}$
        \EndIf
        \State Enumerate $S_i$ to find all connectable motif pairs.
        \State Connect each motif pair to a motif candidate, leading to a candidate set $C_i$.
    \EndFor
    \State Construct a candidate list $\mathcal{C} = C_1 \cup C_2 \cup\cdots\cup C_n$. 
    \State Count frequency $d_i$ of each candidate $c_i \in \mathcal{C}$.
    \State Select the candidate $c_h$ with the highest frequency.
    \State Define $c_h$ as motif and add $c_h$ into $\mathcal{M}$.
    \State Update the representation of each molecule.
    \EndWhile
    \end{algorithmic}
\end{algorithm}

\subsubsection{Data-driven motif decomposition}

\begin{algorithm}
\caption{Data-driven motif decomposition}\label{alg: decomp}
\begin{algorithmic}[1]
\State {\bfseries Input:} Graph $G$, Motif vocabulary $\mathcal{M}$
\State {\bfseries Output:} Motif SMILES list $L$
\State {\bfseries Initialization:} An empty lists $L$
\State $S$, $V$, $E$ $\Leftarrow$ $\textproc{Initialize($G$)}$

\While{True}
    \For{motif pairs in $S$}
        \If{$m$ $\in$ $\mathcal{M}$}
            \State Add $m$ to $L$
            \State Update $S$, $V$, $E$ to merge the motif
        \EndIf
    \EndFor
    \If{no motif selected}
        \State \textbf{break}
    \EndIf
\EndWhile
\end{algorithmic}
\end{algorithm}

In this section, we propose an algorithm that can decompose a given molecule into motifs, using a motif vocabulary as built in~\ref{Task:3.1.2}.

At the start of the process, a motif vocabulary $\mathcal{M}$ is given. We will create an empty motif list to store motifs identified from the molecule. Note that the same motif's multiple instances will all be in the list.
We begin by applying the \textproc{Initialize()} function to obtain three hash maps: $S$, $V$, and $E$. Subsequently, the algorithm enters a greedy process to incrementally identify more complex motifs. It commences from an arbitrary node and generates a connectable motif pair, which is then transformed into a motif candidate. For ease of comparison with motifs in the motif vocabulary, this candidate is converted into a SMILES string representation. A graph-matching process is then carried out against every motif within the motif vocabulary. If the motif candidate is found within the vocabulary, we update the hash maps $S$, $V$, and $E$ to integrate the identified motif. Following this, we proceed to examine the next connectable motif pair.
This process will continue until no further motifs can be identified. Eventually, the algorithm will output a list of motifs decomposed from the molecule. The process is described in Algorithm~\ref{alg: decomp}.

\subsubsection{Time complexity analysis on motif decomposition}
Our proposed algorithm begins with an empty motif list. In the worst-case scenario, where all motif candidates are already present in the motif vocabulary, the algorithm follows a process where a connectable motif pair is merged into a single node during each iteration. In this scenario, the algorithm will iterate a total of $\frac{E}{2}$ times, with $E$ denoting the total number of bonds. Furthermore, in each of these iterations, the algorithm also checks for the existence of the motif candidate within the motif vocabulary.
The motif existing checking algorithm comprises two main operations: the conversion of molecules into SMILES strings and string comparison. Generating canonical
SMILES strings from molecules is fundamentally a depth-first search graph traversal
procedure, which carries a linear computational complexity $O(N + E)$, where $N$ is the number of nodes in the graph. Similarly, the
time complexity for comparing two strings is also linear. Consequently, the overall
time complexity for this graph-matching algorithm is $O(N + E + M)$, where $M$ is the size of the motif vocabulary $\mathcal{M}$. Then, the overall time complexity for the motif decomposition algorithm is $O(E(N + E + M))$

\subsection{Build a Heterogeneous Graph based on MotifPiece Vocabulary}
To fully leverage the motif information, we construct a complex, heterogeneous graph that comprises two distinct node categories: motif nodes and molecule nodes. Each node in the motif category symbolizes a distinct motif, while each node in the molecule category corresponds to a specific molecule. Additionally, this graph incorporates two unique edge types: motif-motif edges and motif-molecule edges.  We establish a connection between two motif nodes with an edge if they share at least one common atom in a molecule. Similarly, we link a motif node with a molecule node if the motif forms part of the molecule. To underscore the relevance of each motif within a molecule, we employ the Term Frequency-Inverse Document Frequency (TF-IDF) algorithm. This method helps us assign properties to the edges between motifs and molecules, effectively highlighting the relative significance of each motif in relation to its specific molecule. To illustrate the relationships among motifs, we use the Pointwise Mutual Information (PMI) algorithm to allocate attributes to the edges between motifs. This allows us to efficiently depict the associations between different motifs.

Our approach differs significantly from the conventional use of GNNs, which typically operate on a single molecular graph. By creating a heterogeneous network structure, we can foster an environment that allows for the transfer of information between diverse motifs and molecules. This information transfer is achieved through a process known as message passing. The message-passing framework enables each node to communicate with its neighboring nodes, facilitating the exchange of crucial information between different motifs and molecules within the graph. Through this unique graph structure and the employment of message passing, we are able to enhance the depth of analysis by interlinking different motifs and molecules, fostering a more holistic understanding of the system being examined. This could lead to uncovering more complex patterns or relationships within the data, further enriching our understanding of the molecule and motif interactions.


To formally describe the process, let's consider a dataset of molecules, denoted as $\mathcal{G} = \{G_1, G_2, ..., G_n\}$, where $n$ represents the total number of molecules within the dataset and $G_i$ symbolizes the $i$th molecule. 
Initially, we employ MotifPiece to extract motifs for each molecule, and we create a motif vocabulary $\mathcal{M} = \{m_1, m_2, ..., m_k\}$, where $k$ signifies the total count of motifs.
In the next step, we create a heterogeneous graph, $H$, which is derived from these motifs. During the inception of this graph, we incorporate all nodes representing both motifs and molecules. A motif node, denoted as $n_m$, signifies a motif graph $m$, while a molecule node, denoted as $n_g$, stands for a molecule graph $G$.
Subsequently, we introduce an edge between two motif nodes $n_m^i, n_m^j$ if $(m_i \vee m_j) \neq \emptyset$. Likewise, we establish a connection between a motif node $n_m^i$ and a molecule node $n_g^k$ if $m_i \subseteq G_k$. At last, we assign edge attributes to two different types of edges following formulas. Given an edge $e_{ij}$ that connect node $i$, and node $j$, we have edge attribute $A_{ij}$
\begin{equation}  
A_{ij} = \left\{  
             \begin{array}{lr}
             \text{PMI}_{ij},        & \text{if $i$, $j$ are motifs} \\
             \text{TF-IDF}_{ij},     & \text{if $i$ or $j$ is a motif} \\
             0,                      & \text{Otherwise}
             \end{array}
\right.
\end{equation}
The TF-IDF value of an edge between a motif node $i$ and a molecular node $j$ is computed as
\begin{equation}
    \text{TF-IDF}_{ij} = C(i)_j \left( \log\frac{1 + N}{1 + N(i)} +1 \right),
\end{equation}
where $C(i)_j$ is the number of times that the motif $i$ appears in the molecule $j$, $N$ is the number of molecules, and $N(i)$ is the number of molecules containing motif $i$.

The PMI value of an edge between two motif nodes is computed as
\begin{equation}
\text{PMI}_{ij} = \log\frac{p(i, j)}{p(i)p(j)},
\end{equation}
where $p(i,j)$ is the probability that a molecule contains both motif $i$ and motif $j$, $p(i)$ is the probability that a molecule contains motif $i$, and $p(j)$ is the probability that a molecule contains motif $j$. We use following formulas to compute these probabilities.
\begin{align}
    p(i,j) = \frac{N(i, j)}{N},  \;\;\;\;
    p(i) = \frac{N(i)}{N}, \;\;\;\;
    p(j) = \frac{N(j)}{N},
\end{align}
where $N(i, j)$ is the number of molecules that contain both motif $i$ and motif $j$. Figure~\ref{fig:hlm} provides an example of heterogeneous motif graph construction. Note that we assign zero weight for motif node pairs with negative PMI value.

\subsection{Heterogeneous Graph Learning Module}\label{sec:3.3}
After the creation of the heterogeneous graph, we employ a Heterogeneous Graph Learning Module for the task of learning the representation of the molecule. This approach effectively capitalizes on the diverse types of nodes and edges within the heterogeneous graph to draw out rich, contextual information about the molecule.

Given a heterogeneous graph $H = \{V, E\}$ and a motif vocabulary $\mathcal{M} = \{m_1, m_2, \ldots, m_k\}$, where $V$ is the node set of the graph and $E$ is the edge set. Let's consider a collection of graphs $\mathcal{G} = \{G_1, G_2, ..., G_i, ..., G_{n}\}$, which comprises all molecule. In this context, each graph $G_i$ from the collection $\mathcal{G}$ uniquely corresponds to a molecule node $v_i$ within the set $V$, while each graph $m_j$ corresponds to a motif node $v_j$ in the set $V$. To ensure precise information extraction at both the atomic and motif levels, we employ two GNNs dedicated to obtaining graph embeddings at the individual node level for motif and molecule graphs. Let $g_1(\cdot)$, $g_2(\cdot)$ denote two GNNs, and 
\begin{align}
    o_i &= g_1\left(X_i, A_i\right) \in \mathbb{R}^{1 \times h_1} \\
    o_j &= g_2\left(X_j, A_j\right) \in \mathbb{R}^{1 \times h_1}
\end{align}
are graph embeddings for a molecule graph $G_i \in \mathcal{G}$ and a motif graph $m_j \in \mathcal{M}$, where $X_i \in \mathbb{R}^{p \times d}$ signifies the node feature matrix of the $G_i$, $A_i \in \mathbb{R}^{p \times p}$ is the corresponding adjacency matrix, $X_j \in \mathbb{R}^{p \times d}$ represents the node feature matrix of the $m_j$, $A_i \in \mathbb{R}^{p \times p}$ is the corresponding adjacency matrix, and $h_1$ is the output dimension of $g_1(\cdot)$ and $g_2(\cdot)$. Then we combine molecule graph embeddings and motif embeddings as $O = \left\{o_1, o_2, ..., o_i, ..., o_{n+k}\right\}$. These graph embeddings play a vital role in capturing and encoding the nearby network structure around nodes. As a result, they are essential for preserving the critical topological information found in the original graph.

We concatenate these learned graph embeddings and then serve as the node features within our heterogeneous graph. They carry a condensed yet information-rich representation of the motifs and molecules from the original node level. We ensure that the individual atomic level feature information and topology information of each motif and molecule are preserved and ready to be processed by the next phase of our module.
Following this, we apply another GNNs to the heterogeneous graph, now armed with the previously learned graph embeddings as node features. Let $g_3(\cdot)$ denote the third GNNs, and
\begin{equation}
    Y' = g_3\left(O, A_H\right) \in \mathbb{R}^{(n+k) \times h_2}
\end{equation}
is the node embeddings of the heterogeneous graph, where $O \in \mathbb{R}^{(n+k) \times h_1}$ is the node features of the heterogeneous graph, $A_H \in \mathbb{R}^{(n+k) \times (n+k)}$ is the corresponding adjacency matrix, and $h_2$ is the output dimension of $g_3(\cdot)$. The third GNNs works to further learn and refine the representation of the graph, taking into consideration the rich, heterogeneous interactions between different types of nodes and edges. The output $Y'$ generated by the second GNN can be applied to a variety of tasks. We then update both GNNs by computing the loss function denoted as
\begin{equation}
\text{loss} = f(Y'[mask], Y),
\end{equation}
where $[mask]$ filters out motif node embeddings, and $Y \in \mathbb{R}^{n \times n}$ is the labels of the dataset. The choice of the loss function $f$ is informed by the requirements of the specific tasks.

This two-step process, involving two GNNs $g_1(\cdot)$ and $g_2(\cdot)$ for node-level graph embedding learning and a subsequent GNN $g_2(\cdot)$ for heterogeneous graph learning, ensures that our model can accurately capture and represent both atom-level and motif-level information. By using this approach, our heterogeneous graph neural network becomes equipped to effectively learn the comprehensive representation of a molecule, encompassing all of its constituent elements and their interactions.

\subsection{Learning on Multiple Molecule Datasets}
The field of molecular biology has seen the emergence of numerous datasets that hold great potential for advanced computational analysis. However, one of the prevalent limitations in the field is the relative smallness of these datasets. Several factors contribute to the scarcity of large molecule datasets. Firstly, the experimental procedures required to generate such datasets are often laborious and time-consuming. Secondly, the highly specialized knowledge required to curate and interpret these datasets restricts their proliferation. Lastly, issues surrounding data privacy and proprietary information often limit the availability of large molecule datasets.

Applying GNNs directly to these small datasets often leads to overfitting. Overfitting is a prevalent problem in machine learning where a model learns the training data too well, capturing noise and outliers in addition to underlying patterns. As a result, such a model performs poorly when faced with new, unseen data, as it fails to generalize effectively. Also, with small datasets, GNNs may not have sufficient data to train effectively, leading to suboptimal performance.

To solve this problem, we leverage the power of the heterogeneous graph to amalgamate diverse datasets with shared motifs. Our approach expands the set of molecule examples in the heterogeneous graph. This expansion aids the model in better generalizing to unfamiliar, unseen data, which is especially beneficial when the constituent datasets exhibit varied distributions. Furthermore, our method helps mitigate the risk of overfitting by enhancing the training data's volume and diversity. This process also fortifies the model against data variations, thereby increasing its versatility and reliability in practical applications. 

Let's take the example of merging two datasets, given two datasets $\mathcal{G}_1 = \{G_1^1, G_2^1, ..., G_i^1, ..., G_n^1\}$ and $\mathcal{G}_2 = \{G_1^2, G_2^2, ..., G_i^2, ..., G_m^2\}$, we apply MotifPiece to extract motif vocabulary $\mathcal{M}_1, \mathcal{M}_2$ for two datasets, respectively. Then, we count the number of common motifs $\mathcal{M}_c = \{m: m \in \mathcal{M}_1 \text{ and } m \in \mathcal{M}_2\}$. If the count of shared motifs, represented by $|\mathcal{M}_c|$, exceeds a given threshold $t$, then we consider merging the two vocabularies into a single vocabulary $\mathcal{M}_C = \{m: m \in \mathcal{M}_1 \text{ or } m \in \mathcal{M}_2\}$. This process allows us to construct a complex heterogeneous graph $H_C$ that accommodates molecule nodes drawn from both datasets.

Similar to section~\ref{sec:3.3}, we start by learning motif and molecule graph embeddings. Given the diverse types of datasets pertaining to different tasks, we opt to employ distinct GNNs to learn from motif graphs and molecule graphs originating from various datasets. For the previous example, we use three GNNs  to learn from motif graphs and molecule graphs in the first and second datasets, respectively. Let $g_1(\cdot)$ be the first GNNs, and
\begin{equation}
    o^m_i = g_1(X^m_i, A^m_i) \in \mathbb{R}^{1 \times h_1}
\end{equation}
represent the graph embedding of a motif graph $G^m_i \in M_C$. The output $O^m = \{o^m_1, o^m_2, ..., o^m_i, ..., o^m_{|M_C|}\}$ is the representation of all motif graphs. Similarly, let $g_2(\cdot)$ and $g_3(\cdot)$ be other two GNNs, we can obtain output $O^1 = \{o^1_1, o^1_2, ..., o^1_i, ..., o^1_n\}$ and $O^2 = \{o^2_1, o^2_2, ..., o^2_i, ..., o^2_m\}$ from two molecule datasets, where
\begin{align}
    o^1_i &= g_2(X^1_i, A^1_i) \in \mathbb{R}^{1 \times h_1}, \\
    o^2_i &= g_3(X^2_i, A^2_i) \in \mathbb{R}^{1 \times h_1}.
\end{align}
Upon obtaining embeddings for motifs and molecules, we concatenate these embeddings to serve as input node features for the heterogeneous graph.
\begin{equation}
    O = O^m \oplus O^1 \oplus O^2,
\end{equation}
where $\oplus$ means concatenation.
Subsequently, we deploy another GNNs to learn the node embeddings of motif and molecule nodes. Let $g_4(\cdot)$ denote the fourth GNNs, and
\begin{equation}
    R = g_4(O, A_{H_C}) \in \mathbb{R}^{(|\mathcal{M}_C|+n+m) \times h_2}
\end{equation}
is the node embedding of the heterogeneous graph.
This is followed by using multiple prediction heads for different datasets to compute the loss. Let $\phi_1(\cdot)$, $\phi_2(\cdot)$ denote the classifier/regressor for two datasets, respectively. We can obtain two losses by
\begin{align}
    l_1 &= f_1\left(\phi_1(R[mask_1]), Y_1\right), \\
    l_2 &= f_2\left(\phi_2(R[mask_2]), Y_2\right),
\end{align}
where $mask_1$ corresponds to the first dataset, while $mask_2$ pertains to the second. Similarly, $Y_1$ and $Y_2$ represent the labels for the first and second datasets, respectively. The loss functions chosen for these two datasets are denoted by $f_1(\cdot)$ and $f_2(\cdot)$.
We aggregate all the losses following the below formula and update all the GNNs concurrently.
\begin{equation}
    Loss = \alpha\cdot l_1 + \beta\cdot l_2,
\end{equation}
where $\alpha$, and $\beta$ serve as the respective weights for the two loss functions.

\subsection{Dataset and settings}
The details of ten datasets are shown in Table~\ref{tab: stat}.

\begin{table}[h]
\caption{Statistics on ten datasets}\label{tab: stat}
\begin{tabular*}{\textwidth}{@{\extracolsep\fill}lccccc}
\toprule
 & PTC\_MR  & PTC\_MM  & PTC\_FR  & PTC\_FM & BBBP \\
\midrule
\# Molecules & 344 & 336 & 351 & 349 & 2039\\
\# Binary prediction tasks & 1 & 1 & 1 & 1 & 1\\
\botrule
\toprule
 & SIDER & CLINTOX & BACE & TOXCAST & TOX21 \\
\midrule
\# Molecules & 1427 & 1478 & 1513 & 8575 & 7831\\
\# Binary prediction tasks & 27 & 2 & 1 & 617 & 12\\
\botrule
\end{tabular*}
\end{table}

Given the datasets originate from two distinct sources, we select evaluation metrics consistent with previous research conventions. For the four datasets sourced from TUDataset, we employ a 10-fold cross-validation approach, given their smaller scale. We adopt validation accuracy as our metric for test accuracy. Due to the limited size of these datasets, test accuracy can fluctuate. Hence, we conduct 10-fold cross-validation ten times, calculating both the mean and standard deviations. Similarly, for the MoleculeNet datasets, in alignment with prior research, we apply a scaffold split method to divide the data into training, validation, and testing sets. Scaffold splitting~\cite{ramsundar2019deep} is commonly perceived as a method that effectively assesses the out-of-distribution generalization capabilities of a model. This approach fundamentally splits datasets based on unique molecular substructures, termed scaffolds, resulting in a rigorous division wherein each subset (training, validation, and testing) contains distinct scaffolds. This characteristic makes it an ideal and challenging test for the robustness and generalizability of a model, as it has to predict on scaffold structures that it has not seen during the training phase. To evaluate the models on these MoleculeNet datasets, we use the Area Under the Receiver Operating Characteristics Curve (AUC-ROC) metric. We run each model ten times and report both the mean and standard deviations to provide a comprehensive performance evaluation.

For all datasets, motif-level embeddings are extracted via a two-layer GIN, and atom-level graph embeddings of each molecule graph are obtained through another two-layer GIN. The node feature of a heterogeneous graph is then formed by concatenating these two embeddings. A further two-layer GIN is used to extract the node embedding of the heterogeneous graph. Lastly, a Multilayer Perceptron (MLP) serves as a classifier to perform the classification task. Each layer employs batch normalization, and dropout is applied to all layers, barring the final layer. In order to ensure a fair comparison, we also implement a four-layer Graph Convolutional Network (GCN) and a four-layer GIN, both with identical batch normalization and dropout settings. The fine-tuned hyperparameters are learning rate and dropout rate. We have considered multiple options for learning rates and dropout probabilities in our experiments. Specifically, learning rates are chosen from the set $\{0.001, 0.002, 0.00002\}$, while the dropout probabilities are selected from the set $\{0.0, 0.5\}$. 

It's worth noting that while our framework primarily utilizes GIN as the underlying backbone, this choice is not restrictive. Indeed, any other GNN variant could be readily substituted into our framework, demonstrating its flexibility and adaptability.

\section{Code availability}
Codes and models are available at
\href{https://github.com/ZhaoningYu1996/MotifPiece}{https://github.com/ZhaoningYu1996/MotifPiece}.

\bibliography{sn-bibliography}
\clearpage

\backmatter

\bmhead{Acknowledgments}
\bmhead{Author Contributions}
\bmhead{Competing Interests}
The authors declare no competing interests.

\bmhead{Data and Code Availability}

\end{document}